\documentclass[conference]{IEEEtran}
\IEEEoverridecommandlockouts
\usepackage{cite}
\usepackage{amsmath,amssymb,amsfonts}
\usepackage{algorithmic}
\usepackage{graphicx}
\usepackage{textcomp}
\usepackage{xcolor}
\usepackage{graphicx}
\usepackage{epstopdf}
\usepackage{amssymb}
\usepackage{amsfonts}
\usepackage{amsmath}
\usepackage{algorithm}
\usepackage{algorithmic}
\usepackage{subeqnarray}
\usepackage{cases}
\usepackage{color}

\usepackage{bm}
\usepackage{subfigure,amsmath,amssymb,cite}
\def\BibTeX{{\rm B\kern-.05em{\sc i\kern-.025em b}\kern-.08em
    T\kern-.1667em\lower.7ex\hbox{E}\kern-.125emX}}
\begin{document}

\title{Deep-learning-aided Low-complexity DOA Estimators for Ultra-Massive MIMO Overlapped Receive Array*\\
\thanks{This work was supported in part by the National Natural Science Foundation of China (Nos.U22A2002, and 62071234), the Major Science and Technology plan of Hainan Province under Grant ZDKJ2021022, and the Scientific Research Fund Project of Hainan University under Grant KYQD(ZR)-21008.}
}


\author{\IEEEauthorblockN{1\textsuperscript{st} Yiwen Chen}
\IEEEauthorblockA{\textit{Department of Information and} \\
\textit{Communication Engineering} \\
\textit{Hainan University}\\
Haikou, Hainan \\
cyw1978650281@163.com}
\and
\IEEEauthorblockN{2\textsuperscript{nd} Yuxiang Zheng}
\IEEEauthorblockA{\textit{Department of Computer Science} \\
\textit{and Technology} \\
\textit{Hainan University}\\
Haikou, Hainan \\
orionzyx@163.com}
\and
\IEEEauthorblockN{3\textsuperscript{rd} Junhao Zhang}
\IEEEauthorblockA{\textit{Department of Information and} \\
\textit{Communication Engineering} \\
\textit{Hainan University}\\
Haikou, Hainan \\
1006577484@qq.com}
\and
\and
\and
\and
\IEEEauthorblockN{4\textsuperscript{th} Xichao Zhan}
\IEEEauthorblockA{\textit{Department of Information and} \\
\textit{Communication Engineering} \\
\textit{Hainan University}\\
Haikou, Hainan \\
17756249373@163.com}
\and
\IEEEauthorblockN{5\textsuperscript{th} Feng Shu}
\IEEEauthorblockA{\textit{Department of Information and} \\
\textit{Communication Engineering} \\
\textit{Hainan University}\\
Haikou, Hainan \\
shufeng0101@163.com}
\and
\IEEEauthorblockN{6\textsuperscript{th} Qijuan Jie}
\IEEEauthorblockA{\textit{Department of Information and} \\
\textit{Communication Engineering} \\
\textit{Hainan University}\\
Haikou, Hainan \\
jieqijuan1012@163.com}
}

\maketitle

\begin{abstract}
Massive multiple input multiple output(MIMO)-based fully-digital receive antenna arrays bring huge amount of complexity to both traditional direction of arrival(DOA) estimation algorithms and neural network training, which is difficult to satisfy  high-precision and low-latency applications in future wireless communications. To address this challenge, two estimators called OPSC and OSAP-CBAM-CNN are proposed in this paper. The computational complexity of the traditional DOA algorithm is first considered to be reduced by dividing the total set of antennas into multiple overlapped subarrays uniformly, each subarray crosses each other proportionally and performs DOA estimation to generate coarse angles, and all angles are coherently combined to get the better estimation, the final DOA estimation can given by maximum likelihood alternating projection(ML-AP) in a very small range, which has a better performance than the direct partitioning of subarrays. To further reduce the complexity of traditional estimation algorithms, deep neural networks(DNN) are utilized to offline train the relationship between the received signal covariance matrix and the estimated angles. Due to the high complexity of the training network based on large-scale arrays, in the OSAP-CBAM-CNN method, the complex network is divided into several smaller networks based on the overlapped subarray to give rough DOA estimations, followed by coherent combining and AP algorithm to get the final DOA estimation.
Simulation results show that as the number of antennas goes to large-scale, the proposed methods can achieve a remarkable complexity reduction over conventional ML-AP algorithm.
\end{abstract}
\section{Introduction}
Target localization is an issue of great interest in wireless communications, and it has been widely used in various modern engineering fields, including assisted driving, navigation, millimeter wave (mmWave) communications,etc\cite{b1}. One of the key part in target localization is the Direction of Arrival (DOA) technology, which provides accurate directional information for beamforming and achieves higher received signal-to-noise(SNR) ratio at the receiver with less transmit power\cite{b2}. In recent years, the application of mmWave technology has enabled more antennas deployed at the same distance, which combined with multiple-input multiple-output (MIMO) receive array to achieve ultra-high angular resolution and accuracy. For the problem of DOA estimation based on massive MIMO, the first and most important thing is to infer the existence of the emitters, which determines whether a DOA estimation is required. Therefore, in \cite{b3}, three high-performance detectors were proposed to accurately infer the existence of passive emitters from the eigen-space of the sample covariance matrix of the received signal.

The problem of direction finding (DF) with narrow-band arrays of sensors have many conventional algorithms. In \cite{b4}, the authors compared the maximum likelihood (ML) estimator and the multiple signal classification (MUSIC) estimator based on subspace decomposition and inferred some properties of Cramer-Rao bound (CRLB), which shows that the ML estimator has better asymptotic performance in coherent signals or low SNR scenarios, but it is requirement of global multidimensional search leads to high complexity. Therefore, in \cite{b5}, the authors have proposed a low-complexity alternating projection (AP) algorithm to compute an accurate DOA estimation of multiple sources in a passive sensor array and verified that the estimator is also applicable to the case of coherent signals, the key of this algorithm is to find more accurate initial values.
As the number of antennas tends to large-scale, the circuit costs and computational complexity of conventional methods based on fully-digital (FD) receive arrays will increase dramatically. Therefore, in \cite{b6}, Shu et al. proposed a low-complexity hybrid analog-digital (HAD) structure with two maximum-receiving-power-based methods to achieve low-complexity and high-precision DOA estimation. To further reduce the time slot required by estimation, a fast ambiguous DOA elimination method was proposed in \cite{b7}, which could find the true emitter direction only using two-time-slot and the author in \cite{b8} have removed phase ambiguity in only single time-slot. The above proposed methods significantly reduce the circuit cost and computational complexity, but accompanied by a substantial performance loss. In order to achieve a significant reduction in complexity while maintaining the high performance of the FD structure, the authors proposed three low-complexity methods for reducing complexity of eigenvalue decomposition in \cite{b9}, the latter two methods could achieve more than two orders of magnitude complexity reduction and achieve excellent performance.

With the rise of applications such as metaverse and web 3.0, it is urgent to satisfy the need for high transmission rates and low latency. Machine learning and deep learning (DL) based DOA estimation can address this problem well. In \cite{b10}, Zhuang et al. formulated a machine learning framework to improve the estimation accuracy of DOA and proposed three weight combiners to combine the maximum likelihood learning output of the training dataset and the real-time estimation dataset. In \cite{b11}, the authors combined DL, uniform circular array (UCA), and HAD to propose a low-complexity estimator. In \cite{b12}, convolutional neural networks (CNN) was utilized to extract features and the model was treated as a multi-label classification problem for prediction. This method achieves better performance than traditional methods at low SNR, but it has substantial performance loss at high SNR due to the grid limitation. In \cite{b13}, the authors analyzed the performance loss of DOA estimation using a receiver array with a low-resolution analog-to-digital converter (ADC) and found a good balance between performance and circuit cost.

In general, as the antenna tends to large-scale, it inevitably brings huge/ultra huge complexity to the traditional DOA algorithms and neural network training. In particular, the ML based DOA estimation algorithm has very good asymptotic performance than other algorithm in scenarios such as coherent sources, low SNR, etc., but its requirement of global multidimensional search and smaller grid causes very high complexity as the antenna tends to large-scale.
\begin{itemize}
\item In order to reduce the computational complexity of conventional algorithm based on massive MIMO, the overlapped partitioned subarray coherent combining (OPSC) method is proposed to address this problem. The total antennas are uniformly divided into multiple subarrays, and each subarray is crossed in a predetermined proportion, the coarse DOA estimations are estimated separately by these subarrays and a more accurate estimation is given by coherent combining, followed by ML-AP in a very small range to get the final estimation. It can be seen that the OPSC method can significantly reduce the computational complexity and achieve a better performance. However, when the Tera-hertz technology brings the ultra large-scale antennas to the public, which long estimation time is difficult to be satisfied for some applications like metaverse and web3.0.
\item The DOA estimation method based on DL can solve the above problem efficiently, but at the same time, as the number of antennas tends to ultra large-scale and the strict requirements for prediction performance, the complexity of DNN training has become a major problem, which is the main reason why few scholars apply DL to the DOA estimation of massive MIMO. Therefore, the OSAP-CBAM-CNN method is proposed to exploit DL to balance the complexity between training and estimation. The signal covariance matrix of each overlapped subarray is fed into the DNN separately, and the complex network is divided into several small networks for training to give the rough DOA estimations and through coherent combining to get the more accurate estimation. The final DOA estimation is given by ML-AP in a very small area. The OSAP-CBAM-CNN method makes a significant complexity reduction while achieving the CRLB.

\end{itemize}

\section{System model}
Assuming $Q$ emitters signal from the direction $\theta= [\theta_1,\theta_2\cdots,\theta_Q]$ impinges on the uniformly-spaced linear array (ULA) with $N$ antenna elements, each antenna gets a different version of the received signal due to the propagation delay and the received baseband signal vector is expressed as
\begin{align}\label{y}
\mathbf{y}(n)=\mathbf{A}(\theta)\mathbf{s}(n)+\mathbf{v}(n)
\end{align}
where $\mathbf{v}(n)\sim\mathcal{C}\mathcal{N}(0,\sigma^2_w\textbf{I}_N)$ is the additive white Gaussian noise (AWGN) vector, $\mathbf{s}(n)=[s_1(n),\cdots,s_Q(n)]^T$ is the Q emitter signals, $ \mathbf{A}(\theta)=[\mathbf{a}(\theta_1),\cdots,\mathbf{a}(\theta_Q)] $ is the array manifold and $\mathbf{a}(\theta_q)$ is defined as
$
\textbf{a}(\theta_q)=[1,e^{j\frac{2\pi d\sin\theta_q}{\lambda}},\cdots,e^{j\frac{2\pi(N-1)d\sin\theta_q}{\lambda}}]^T,
$
where $\lambda$ is the wavelength of the carrier frequency, and $d=\frac{\lambda}{2}$. Here, the phase reference point is chosen to be the left of the array.
\section{Conventional ML Estimator and AP algorithm}
DOA estimation based on ML methods has very good asymptotic performance compared to the super-resolution algorithm based on subspace decomposition in the coherent source and single snapshot scenarios, which function can be given by
\begin{align}\label{ML}
\mathbf{\hat{\theta}}=\mathop{\mathrm{argmax}}\limits_{\mathbf{\hat{\theta}}\in\mathbf{\Psi}}{\textmd{tr}\{P_{\mathbf{A}}\hat{\mathbf{R}}\}}.
\end{align}
where $\hat{\mathbf{R}}$ is the sample covariance matrix, $P_{\mathbf{A}({\theta})}=\mathbf{A}({\theta})\left(\mathbf{A}^H({\theta}) \mathbf{A}({\theta})\right)^{-1} \mathbf{A}^H({\theta})$ and $\mathbf{\Psi}=\{(\theta_1,\cdots,\theta_q):\theta_1\in[-\pi/2:\sigma:\pi/2],\cdots,\theta_q\in[-\pi/2:\sigma:\pi/2]\}$, $\sigma$ is the search grid and the point of $\mathbf{\Psi}$ is $\epsilon=(\pi/\sigma+1)^q$. It can be seen that the need for multi-dimensional search and small grid make ML estimator difficult to exploited. Therefore, the AP algorithm combines alternating optimization methods and projection matrix decomposition methods to transform the multi-dimensional search into multiple single-dimensional searches, which significantly reduce the complexity of ML method. Consequently, the estimation of the $ith$ iteration of the AP algorithm is given by
\begin{align}\label{AP}
\hat{\theta}_q^{(i)}=\max _{\theta_q}\left\{\mathbf{b}^{\mathrm{H}}\left(\theta_q, \mathbf{\Theta}_{(q)}^{(i-1)}\right) \hat{\mathbf{R}} \mathbf{b}\left(\theta_q, \mathbf{\Theta}_{(q)}^{(i)}\right)\right\}
\end{align}
where $\mathbf{\Theta}_{(q)}^{(i)}=[\hat{\theta}_{(1)}^{(i)},\cdots,\hat{\theta}_{(q-1)}^{(i)},\hat{\theta}_{(q+1)}^{(i)},\cdots,\hat{\theta}_{(Q)}^{(i)} ]$ and $\mathbf{b}(\theta_q, \mathbf{\Theta}_{(q)}^{(i)})$ is a unit vector defined as
\begin{align}\label{b}
\mathbf{b}\left(\theta_q, \mathbf{\Theta}_{(q)}^{(i)}\right)=\mathbf{I}-\mathbf{P}_{A\left(\hat{\mathbf{\theta}}_{(q)}^{(i)}\right)}\left\|\mathbf{I}-\mathbf{P}_{A\left(\hat{\mathbf{\theta}}_{(q)}^{(i)}\right)}\right\|^{-1}
\end{align}
where $\mathbf{I}$ is a identity matrix and $\|\cdot\|$ denotes the norm. AP algorithm requires iterations to get the final DOA estimation, and its crucial to find an excellent initial value.
\section{Proposed two low-complexity structures and estimators}
To address the problem that the complexity of ML-AP algorithm and the neural network training is very high due to the missive MIMO array. A low-complexity estimation method is proposed firstly, and a CNN based method is exploited on the basis of this overlapped structure to further reduce the complexity and balance the training with the estimation.
\subsection{Proposed OPSC}
\begin{figure}[h]
\centering
\includegraphics[width=0.30\textwidth]{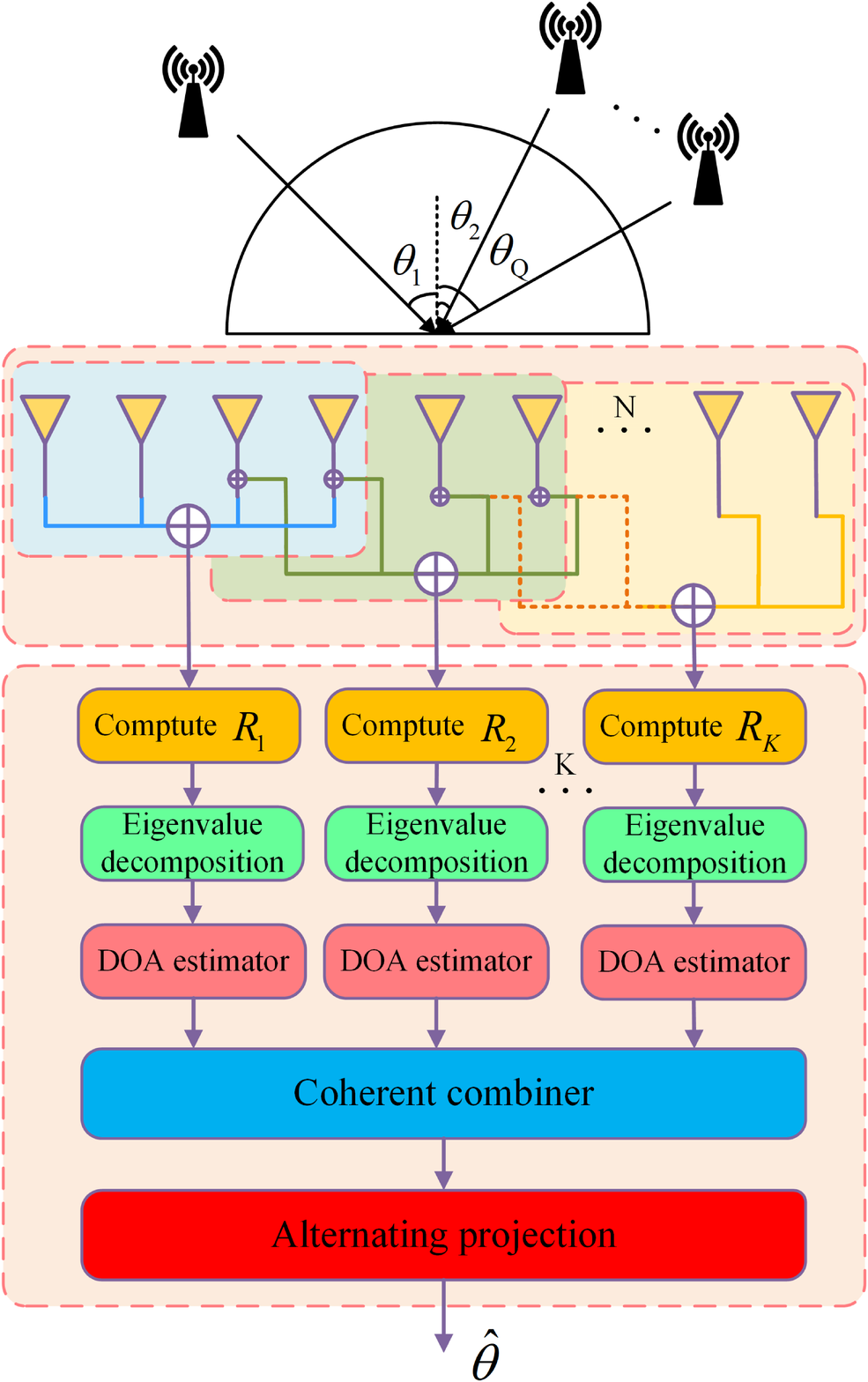}\\
\caption{Proposed a low-complexity  overlapped partitioned subarray combining structure}\label{system_model.eps}
\end{figure}
The Fig.~\ref{system_model.eps} shows the structure of proposed OPSC estimator. $N$ antennas are divided equally into $ K $ subarrays with each subarray containing $M$ antennas, and two adjacent sub-arrays cross each other in proportion $M_0=\alpha M$,where $\alpha\in[0,1]$. Therefore, we can calculate $K=(N-M_0)/(M-M_0)$. It can be seen that, compared to no overlapped partitioned subarray(NOPS), the OPSC can improve the estimation performance by changing the number of crossed antennas and not changing the number of antennas in subarray, which is certainly effective in reducing the computational complexity. After the parallel RF chain and ADC, the receive signal vector of $kth$ overlapped subarray is given by
\begin{align}
\textbf{y}_k(n)=\textbf{A}_k(\theta)\textbf{s}_{k}(n)+\textbf{v}_k(n),k=1,2,...,K
\end{align}

where $\textbf{A}_k(\theta)=[\textbf{a}_k(\theta_1),\cdots,\textbf{a}_k(\theta_Q)]$ is the array manifold of the kth subarray and $\textbf{a}_k(\theta_q)$ defined as
$
\textbf{a}_k(\theta_q)=[e^{j2\pi\frac{(kM-M-kM_0+M_0)d\sin\theta_q}{\lambda}},\cdots,e^{j2\pi\frac{(kM-kM_0+M_0-1)d\sin\theta_q}{\lambda}}]^T,
$.
Assuming the signal and noise are uncorrelated, the covariance matrix of received signal is expressed as
$\mathbf{R}_k =\mathrm{E}\left[\mathbf{y}_k(n) \mathbf{y}_k^H(n)\right]$.
In practical applications, the covariance matrix cannot be got directly. Therefore, the sample covariance matrix and the corresponding eigen-decomposition is given by
\begin{align}\label{eg}
\textbf{R}_{k}=\frac{1}{J}\sum^J_{l=1}\textbf{y}_k(j)\textbf{y}^H_k(j)=[\textbf{U}_S\,\textbf{U}_N]\Sigma[\textbf{U}_S\,\textbf{U}_N]^H
\end{align}
where $J$ is the snapshots, $ \textbf{U}_S $  and $ \textbf{U}_N $ stand for the signal and noise subspaces, respectively. Based on the above eigenvalue decomposition and the Root-MUSIC method, the corresponding estimated angle can calculated as follows
\begin{align}\label{rd}
\hat{\theta}_k=\arcsin \left[\frac{\lambda}{2 \pi d} \arg \left(\hat{z}_q\right)\right], \quad q=1, \cdots, Q
\end{align}
where $\hat{z}_q$ is the $Q$ roots of the largest magnitude of the polynomial, the detailed information of Root-MUSIC can be find here\cite{b14}.
A more accurate DOA estimation can be given by the coherently combining of $K$ overlapped subarrays as follows
\begin{align}\label{cc}
\hat{\theta}=w_1\theta_1+w_2\theta_2+\cdots+w_k\theta_K,k=1,2,...,K
\end{align}
where $w_k$ is the weight parameter and are taken to be $1/K$ in here.
Construct the candidate-angle-set with $\hat{\theta}$ as the center can be expressed as $\tilde{\mathbf{\Psi}}=\{(\theta1,\cdots,\theta_q):\theta_1\in[\hat{\theta}_1-h\sigma:\sigma/p:\hat{\theta}_1+h\sigma],\cdots,\theta_q\in[\hat{\theta}_q-h\sigma:\sigma:\hat{\theta}_q+h\sigma]\}$
, where $h$ and $p$ are arbitrary constants.
Due to the application of the AP algorithm, the point of set is $\tilde{\epsilon}=Q(2hp+1)$.
Finally, substitute $\hat{\theta}$ into (\ref{AP}) as the initial value for a small range of AP iterations to get the perfect DOA estimation $\tilde{\mathbf{\theta}}$.

Consequently, the proposed OPSC method can significantly reduce the complexity by adjusting the number of subarray antennas $M$ or the number of crossed antennas $M_0$, which is able to reach convergence with a very small number of iterations. However, it is still difficult to be utilized in the face of future speed-critical applications like metaverse. Therefore, the method based on CNN is proposed to address the dilemma.

\subsection{Proposed OSAP-CBAM-CNN}
\begin{figure}[h]
\centering
\includegraphics[width=0.40\textwidth]{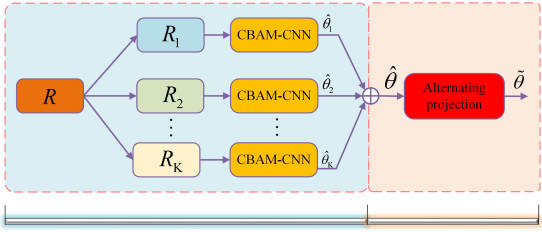}\\
\caption{Framework of proposed OSAP-CBAM-CNN structure}\label{figure2_PI.eps}
\end{figure}
Neural network-based DOA estimation provides high performance, but very few scholars have used it in large-scale MIMO scenarios  mainly because of the extremely high complexity of the training process. To further address the dilemma and give a high-performance solution, the structure of the proposed OSAP-CBAM-CNN method with $N$ antennas is shown in Fig. ~\ref{figure2_PI.eps}. In this structure, $K$ small networks are utilized instead of the large network based on the overlapped structure and let us define a $M\times M\times 3$ real vector $\mathbf{\hat{X}}$ as
\begin{align}
\mathbf{\hat{X}}=\left\{
\begin{aligned}
\mathbf{X}_{:,:,1} & = & \textmd{Re} \{\textbf{R}_k\} \\
\mathbf{X}_{:,:,2} & = & \textmd{Im} \{\textbf{R}_k\} \\
\end{aligned}
\right.
\end{align}
where 
$\textmd{Re}\{\mathbf{R}_k\}$ and $\textmd{Im}\{\mathbf{R}_k\}$ denote the real and the imaginary part of $\mathbf{R}_k$, respectively. Assuming DOA prediction is modeled as a regression task and the training data angles are chosen from $-\theta$ to $\theta$ grid of $g$, with a total of $\phi=2\theta/g+1$ angles. For each SNR level, there are $Z=C_\phi^Q$ covariance matrix data when the emission sources are $Q$, where $C_.^.$ is the permutation and combination, each covariance matrix corresponds to $Q$ angles,
which gives the process of data management and labeling.
It can be found that $\hat{\mathbf{X}}$ is a function of $\theta$.
Therefore, we are going to introduce the architecture of proposed NN.

The proposed neural network is first a $C$-layer convolutional network, each layer consists of a 2D convolutional layer with $F$ filters, a batch normalization layer and an activation layer, the nonlinear activation function used  here is the \verb"ReLU", which can be expressed as $\verb"ReLU" =max(0,x)$. The convolution kernel of $\emph{e}_c\times \emph{e}_c$ is used to extract features by stride $\emph{S}$. Thus, assume the input dimension is $M\times M\times 2$, the output dimension after the first convolutional layer can be expressed as $M_1\times M_1\times F$, where $M_1=(M-\emph{e}_1)/$\emph{S}$+1$, and through the $C$ convolutional layer, the output feature dimension can be further represented as $\mathbf{O_C}:M_c\times M_c\times F$. Then, a convolutional block attention module (CBAM) is added to optimize the neural network structure, which consists of a channel attention module (CAM) and a spatial attention module (SAM). The former is used to show the correlation between different channels, and the importance of each feature channel is automatically obtained through network learning, and the feature map through CAM is given by
$
\mathbf{A}_{\mathbf{c}}(\mathbf{O_C})=\sigma(MLP(AP(\mathbf{O_C}))+M L P(MP(\mathbf{O_C})))
$,
where MLP is a multiple perception network, AP is average pooling and MP is max-pooling.
The SAM is used to enhance the feature representation of key regions, and the output feature map can be represented as
$
\mathbf{A}_{\mathbf{S}}(\mathbf{O_C})=\sigma(f^{7\times 7}(AP(\mathbf{O_C});MP(\mathbf{O_C})))
$.
CBAM combines CAM and SAM to achieve a sequential attention structure from channel to space, and the final encoder output can be given by
$
\mathbf{A}=\mathbf{A}_{\mathbf{S}}(\mathbf{A}_{\mathbf{c}}(\mathbf{O_C})\otimes\mathbf{O_C})\otimes(\mathbf{A}_{\mathbf{c}}(\mathbf{O_C})\otimes\mathbf{O_C})
$.
It should be noted that due to the low complexity of the internal structure of CBAM, it is well suited to be used in the neural network proposed in our manuscript.

In contrast to the extracted feature part of the network, $D$-layer fully connected(FC) layers are used to restore the extracted features to the form of the original input. Each FC layer consists of a dense layer of $n_{ds}$ neurons, a $20\%$ Dropout layer, and 
we use a regression model based on the proposed CNN structure instead of classification model, with the last layer consisting of a dense layer containing $Q$ neuron to map the angle directly, which can achieve better performance at any SNR than the classification model.

Since the proposed neural network is a regression problem, in order to measure the loss of it, the loss of each angle of corresponding $\hat{\mathbf{X}}$ are calculated and averaged, The MSE loss function used here is given by
\begin{align}
\mathcal{L}(\theta,\tilde{\theta})=\frac{1}{Q}\sum_{i=1}^Q(\theta_i-\tilde{\theta}_i)^2
\end{align}
Through the designed network, the mapping of inputs to outputs can be expressed as
\begin{align}
\hat{\mathbf{\theta}}_k=f_D(\cdots f_{D_1}(f_{C}(\cdots(f_{C_1}(\mathbf{\hat{X}}))))
\end{align}
By coherently combining the estimation of $K$ small networks, a more accurate DOA estimate can be given as
\begin{align}
\hat{\mathbf{\theta}}=\frac{1}{K}\sum_{k=1}^K\hat{\mathbf{\theta}}_k
\end{align}
Slimily to (\ref{cc}), the final DOA estimation can be calculated by AP algorithm.
\subsection{Complexity Analysis}
Below, we make an analysis of computational complexities of the proposed estimators with traditional ML-AP algorithm as a complexity reference. Thus, the complexity of OPSC is as follows $C_{OPSC}=O\{K\left(M^3-M^2+ML(2M+1)\right)\}$
FLOPs. And the complexity of conventional ML-AP is $ C_{ML-AP}=O\{(\pi/\sigma+1)q(2N^2(Q-1)+3N^2+4N(Q-1)^2)\} $
From the analysis above, it can be found that the complexity of OSAP-CBAM-CNN comes from the calculation of $\hat{\mathbf{\theta}}_k$ and the iterative optimal estimation $\tilde{\mathbf{\theta}}$.

The complexity of calculating $\hat{\mathbf{\theta}}_k$ is related to the structure of the neural network (i.e., the number of layers and the parameters in each layer) and it can be given by $C_{NN}=O\{\sum_{k}^K\sum_{c=1}^C(M_{c}^2)^{(k)}(\emph{e}_{c}^2)^{(k)}(F_{c-1}^2)^{(k)}(F_{c}^2)^{(k)}+\sum_{k}^K\sum_{ds=1}^Dn_{ds}^kn_{ds+1}^k
\}$. Since all networks are not large and only operated between real numbers when computing $\hat{\mathbf{\theta}}_k$. Therefore, when the number of antennas tends to large-scale, the complexity of the proposed estimation method will be mainly reflected in the selection of the final estimation $\tilde{\mathbf{\theta}}$, which will be about
$C_{OSAP-CBAM-CNN}=O\{\tilde{\epsilon}(2N^2(Q-1)+3N^2+4N(Q-1)^2)
\}$. Compared with the FD ML-AP estimator, the computational complexity of the proposed two estimators is significantly reduced, especially as the number of antennas tends to large-scale.

\section{Simulation Results}
In this section, we present simulation results to assess the performance of proposed DOA estimators: OPSC and OSAP-CBAM-CNN with FD ML-AP\cite{b5} as reference. Assuming two source impinge on the array and $L = 1000$, the training signal source is
within the angular range $[-60^o,60^o]$.
\begin{figure}[h]
\centering
\includegraphics[width=0.38\textwidth]{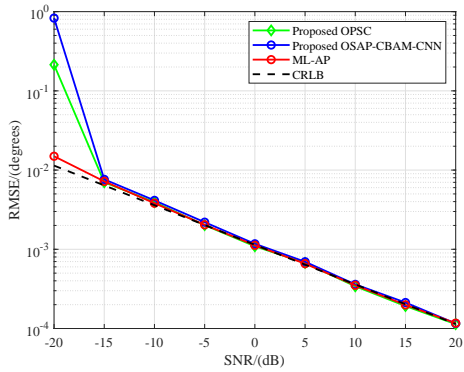}\\
\caption{RMSE versus SNR of the proposed method}\label{figure3_RMSE.eps}
\end{figure}

Fig. \ref{figure3_RMSE.eps} plots the root mean squared error (RMSE) versus SNR of the two proposed DOA estimators OPSC and OSAP-CBAM-CNN for $ N=128 $, $ M=32 $ and $ M_0=16 $, where the corresponding CRLB is used as a performance benchmark. From Fig. \ref{figure3_RMSE.eps}, it is seen that the proposed OSAP-CBAM-CNN method and OPSC method both can achieve the corresponding CRLB when SNR exceed -15dB.

\begin{figure}[h]
\centering
\includegraphics[width=0.38\textwidth]{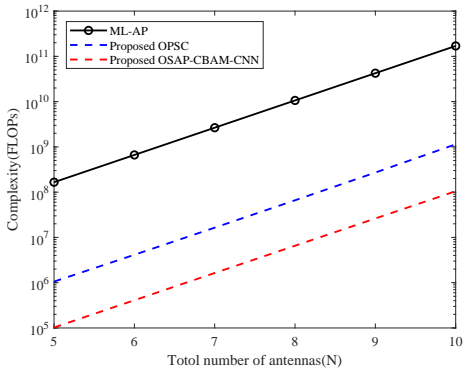}\\
\caption{Complexity versus $ \log_2N $($N$ is the number of antennas) }\label{figure5_complexity.eps}
\end{figure}

Fig. \ref{figure5_complexity.eps} shows the computational complexities versus the number of antennas with $M_0=1/2M=1/8N$ and $ N $ varying from 32 to 1024. From this figure, it is seen that as the number of antennas goes to ultra-large-scale, the computational complexities of our proposed two methods are two-order-of-magnitude to three-order-of-magnitude lower in terms of FLOPs than conventional ML-AP algorithm.
\section{Conclusions}
In this paper, based on the large-scale MIMO receive array, two low-complexity DOA estimators are proposed: OPSC and OSAP-CBAM-CNN. As the number of antennas tends to large-scale, the proposed two methods have more than two orders of magnitude of complexity reduction compared to the traditional ML-AP method and achieve excellent estimation performance. In particular, the OSAP-CBAM-CNN method combines DNN with traditional algorithm and partitioning the complex network into multiple small networks based on overlapped subarray, which not only improves the estimation performance but also significantly reduces the computational complexity. The two proposed low-complexity algorithms will help DOA estimation for massive MIMO to satisfy the low-latency requirements of future applications like beyond 5G/6G.

\vspace{12pt}

\end{document}